\documentclass[conference]{IEEEtran}

\usepackage{amsmath}
\usepackage{amsmath,amsthm}
\usepackage{cite}   % Written by Donald Arseneau
\usepackage{graphicx}  % Written by David Carlisle and Sebastian Rahtz
\usepackage{amsmath, amsopn, psfrag, amsthm}   % From the American Mathematical Society
\usepackage{amssymb}
\usepackage{color}

\usepackage{algorithm}

\usepackage{algorithmic}
\usepackage[utf8]{inputenc}
\usepackage[english]{babel}
\newtheorem{theorem}{Theorem}
\usepackage{psfrag}
\usepackage{epsfig}
\graphicspath{{./img/}}
 \usepackage{graphicx}
\DeclareGraphicsExtensions{.jpeg,.png, .eps, .pdf}
\usepackage{amssymb,amsmath,mathtools}

\def\bgama{{\boldsymbol{\Gamma}}}
\def\bPhi{{\boldsymbol{\Phi}}}

\def\bdel{{\boldsymbol{\Delta}}}

\usepackage{cite}   % Written by Donald Arseneau
\usepackage{graphicx}  % Written by David Carlisle and Sebastian Rahtz
\usepackage{amsmath, amsopn}   % From the American Mathematical Society
\usepackage{amssymb}
\usepackage{color}
\usepackage{psfrag}
\usepackage{epsfig}
\usepackage{stfloats}
\usepackage{lipsum}
\usepackage{multicol}
\graphicspath{{./img/}}
\DeclareGraphicsExtensions{.jpeg,.png}
\begin{document}
\setlength{\columnsep}{.261in}

\title{Enhanced Max-Min SINR for Uplink Cell-Free
Massive MIMO Systems}
\vspace{-0.1in}
\linespread{.96}
\author{
\IEEEauthorblockN{Manijeh Bashar\IEEEauthorrefmark{1}, Kanapathippillai Cumanan\IEEEauthorrefmark{1}, Alister G. Burr\IEEEauthorrefmark{1}, Mérouane Debbah \IEEEauthorrefmark{2}, and Hien Quoc Ngo\IEEEauthorrefmark{3}}
\IEEEauthorblockA{\IEEEauthorrefmark{1}Department of Electronic Engineering, University of York, Heslington, York, UK}, \IEEEauthorblockA{\IEEEauthorrefmark{2}Large Networks and Systems Group, CentraleSupélec,
Université Paris-Saclay, France},\IEEEauthorblockA{\IEEEauthorrefmark{3}School of Electronics, Electrical Engineering and Computer Science, Queen's University Belfast, Belfast, UK}
Email:{ \{mb1465, kanapathippillai.cumanan, alister.burr\}@york.ac.uk}, m.debbah@centralesupelec.fr, hien.ngo@qub.ac.uk 
}
%Boulogne-Billancourt 92100, France}
%\author{Manijeh Bashar, Kanapathippillai Cumanan,~\IEEEmembership{Member,~IEEE}, Alister G. Burr,~\IEEEmembership{Member,~IEEE}, and Mérouane Debbah,~\IEEEmembership{Fellow,~IEEE}\thanks{M. Bashar, K. Cumanan and A. G. Burr and  are with the Department of Electronic Engineering, University of York, Heslington, York, UK. e-mail: \{mb1465, kanapathippillai.cumanan, alister.burr\}@york.ac.uk.} \thanks{M. Debbah is with the Large Systems and Networks Group, CentraleSupélec,
%Université Paris-Saclay, Gif-sur-Yvette 91192, France, and also with the
%Mathematical and Algorithmic Sciences Lab, Huawei Technologies Co., Ltd.,
%Boulogne-Billancourt 92100, France (e-mail: m.debbah@centralesupelec.fr, {merouane.debbah@huawei.com)}.}
%}
\maketitle
\begin{abstract}
In this paper, we consider the max-min signal-to-interference plus noise ratio (SINR) problem for the uplink transmission of a cell-free Massive multiple-input multiple-output (MIMO) system.
Assuming that the central processing unit (CPU) and the users exploit only the knowledge of the channel statistics, we first derive a closed-form expression for uplink rate.
In particular, we enhance (or maximize) user fairness by solving the max-min optimization problem for user rate, by power allocation and choice of receiver coefficients, where the minimum uplink rate of the users is maximized with available transmit power at the particular user. Based on the derived closed-form expression for the uplink rate, we formulate the original user max-min problem to design the optimal receiver coefficients and user power allocations. However, this max-min SINR problem is not jointly convex in terms of design variables and therefore we decompose this original problem into two sub-problems, namely, receiver coefficient design and user power allocation.
By iteratively solving these sub-problems, we develop an iterative algorithm to obtain the optimal receiver coefficient and user power allocations. In particular, the receiver coefficients design for a fixed user power allocation is formulated as generalized eigenvalue problem whereas a geometric programming (GP) approach is utilized to solve the power allocation problem for a given set of receiver coefficients.
Numerical results confirm a three-fold increase in system rate over existing schemes in the literature.
\vspace{.05cm}

\end{abstract}
% MIMO 5g AP bs ap comp cpu tdd  fdd csit mmse csi csit sinr mrc gp snr mrt soc
\section{Introduction}
 \let\thefootnote\relax\footnotetext{The work of K. Cumanan and A. G. Burr was supported by H2020- MSCA-RISE-2015 under grant number 690750. The work on which this paper is based was carried out in collaboration with COST Action CA15104 (IRACON).}
Massive multiple-input multiple-output (MIMO) is one of the most promising techniques for 5th Generation (5G) networks due to its potential for significant rate enhancement and spectral as well as energy efficiency \cite{5gdebbah,multidebbah,ourvtc18,ouricc2,ouriet_mic,slock_fromMU}. In cell-free Massive MIMO, randomly distributed access points (APs) jointly serve distributed users.
In this paper, we propose a max-min signal-to-interference plus noise ratio (SINR) approach for an uplink cell-free Massive MIMO system. In \cite{debbahmaxmin}, the authors investigate the problem of max–min SINR in a single-cell Massive MIMO system. A similar max-min SINR problem refereed to SINR balancing has been considered for cognitive radio network in \cite{cuma_jointbf_twc10, cuma_sinr_spl10,cuma_sinr_vtc9,cuma_wcnc13,cuma_rate_icc11}. In \cite{marzetta_free16}, the same max-min SINR problem is considered through appropriate user power allocation where a bisection search method is utilized to determine the optimal solution. However, a novel approach to significantly improve all users' performance is proposed in this paper by designing optimal receiver coefficients and user power allocation. By employing maximal ratio combining (MRC) at the receiver, we first derive a closed-form expression for the average uplink rate of the users. Based on these user power allocations and receiver coefficients, we formulate the corresponding max-min SINR problem, which is not jointly convex in terms of the design parameters. In order to realize a solution for this non-convex problem, we decompose the original problem into two sub-problems: receiver coefficient design and user power allocation. An iterative algorithm is
proposed that successively solves these two sub-problems while one of the design variables (i.e., user power allocation or receiver coefficients) is fixed. The receiver coefficient design is formulated into a generalized eigenvalue problem \cite{bookematrix} whereas a geometric programming (GP) approach \cite{bookboyd} is exploited to solve the user power allocation problem. The performance of the proposed scheme in terms of the user rate is significantly higher than that of the scheme proposed in \cite{marzetta_free16}.
The contributions and the results of our work are summarized as follows:
\textbf{1)} For the considered cell-free Massive MIMO system, we derive the average user rate in the uplink.
\textbf{2)} Based on the derived user rate, we propose a novel max-min SINR approach to significantly improve the SINR performance in terms of the achieved user rate. The original max-min problem formulation is not convex and therefore we decompose the original problem into two sub-problems and propose an iterative algorithm to yield the optimal solution.
\textbf{3)} The user power allocation and the receiver coefficient design sub-problems are solved through the GP approach and the generalized eigenvalue problem, respectively.
\textbf{4)} Numerical results are provided to validate the superiority of the proposed algorithm in comparison with the scheme proposed in \cite{marzetta_free16}.
%
%\textit{Outline:} The reminder of the paper is organized as follows. Section II describes
%the system model. Section III investigates the corresponding performance analysis. The proposed max-min SINR approach is presented in Section IV. Numerical results are provided in Section V. Finally, Section VI concludes the paper.
\begin{figure}[t!]
\center
\includegraphics[width=64mm]{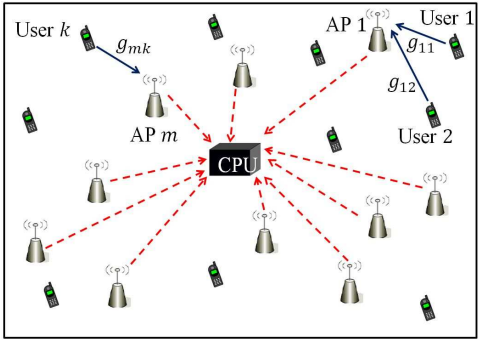}
\vspace{-0.05in}
\linespread{.98}
\caption{The uplink of a cell-free Massive MIMO system with \textit{K} users and \textit{M} APs. The dashed lines denote the uplink channels and the solid lines present the backhaul links from the APs to the central processing unit (CPU).}
\label{sysmodel}
\end{figure}
\section{SYSTEM MODEL}
We consider uplink transmission in a cell-free Massive MIMO system with $M$ randomly distributed single-antenna APs and $K$ randomly distributed single-antenna users in the area, as shown in Fig. \ref{sysmodel}. The channel coefficient between the $k$th user and the $m$th AP,  $g_{mk}$, is modeled as \cite{marzetta_free16}
$
g_{mk}=\sqrt{\beta_{mk}}h_{mk},
\label{g}
$
where $\beta_{mk}$ denotes the large-scale fading and $h_{mk}\sim  \mathcal{CN}(0,1)$ represents small-scale fading between the $k$th user and the $m$th AP.
\subsection{Uplink Channel Estimation}
In order to estimate the channel coefficients in the uplink, the APs employ an minimum mean square error (MMSE) estimator. All pilot sequences used in the channel estimation phase are collected in a matrix $\bPhi \in \mathbb{C}^{\tau\times K}$, where $\tau$ is the length of the pilot sequence for each user and the $k$th column, $\pmb{\phi}_k$, represents the pilot sequence used for the \textit{k}{th} user. After performing a de-spreading operation, the MMSE estimate of the channel coefficient between the $k$th user and the $m$th AP is given by \cite{marzetta_free16}
\begin{IEEEeqnarray}{rCl}
\small
\!\!\!\!\hat{g}_{mk}\!=\!c_{mk}\!\!\left(\!\!\sqrt{\tau p_p}g_{mk}\!+\!\sqrt{\tau p_p}\sum_{k^\prime\ne k}^{K}g_{mk^\prime}\pmb{\phi}_k^H\pmb{\phi}_{k^\prime}\!+\!\pmb{\phi}_k^H\textbf{n}_{p,m}\!\!\right)\!\!\!,~
\label{ghat}
\end{IEEEeqnarray}
where each element of $\textbf{n}_{p,m}$, $n_{p,m} \sim  \mathcal{CN}(0,1)$, denotes the noise at the $m$th antenna, $p_p$ represents the normalized signal-to-noise ratio (SNR) of each pilot sequence (which we define in Section V), and $c_{mk}$ is given by
$
c _{mk}=\dfrac{\sqrt{\tau p_p}\beta_{mk}}{\tau p_p\sum_{k^\prime=1}^{K}\beta_{mk^\prime}\left |\pmb{\phi}_k^H{\pmb{\phi}}_{k^\prime}\right |^2+1}.
\label{cmk}
$
The estimated channels in (\ref{ghat}) are used by the APs to design the receiver coefficients and determine power allocations at users to maintain user fairness. In this paper, we investigate the cases of both random pilot assignment and orthogonal pilots in cell-free Massive MIMO. Here the term "orthogonal pilots" refers to the case where unique orthogonal pilots are assigned to all users, while in "random pilot assignment" each user is randomly assigned a pilot sequence from a set of orthogonal sequences of length $\tau$ ($<K$), following the approach of \cite{marzetta_free16,pilotahmadi}.
\subsection{Uplink Data Transmission}
In this subsection, we consider the uplink data transmission, where all users send their signals to the APs. The transmitted signal from the $k$th user is represented by
$
x_k= \sqrt{q_k}s_k,
$
where $s_k$ ($\mathbb{E}\{|s_{k}|^2\} = 1$) and $q_k$ denote respectively the transmitted symbol and the transmit power at the \textit{k}th user. The received signal at the $m$th AP from all users is given by
$
y_m= \sqrt{\rho}\sum_{k=1}^{K}g_{mk}\sqrt{q_k}s_k+n_m,
$
where $n_m\sim \mathcal{CN}(0,1)$ is the noise at the $m$th AP. In addition, MRC is employed at the APs. More precisely, the received signal at the $m$th AP, $y_m$, is first multiplied with $\hat{g}_{mk}^\ast$. The resulting $\hat{g}_{mk}^\ast y_m$ is then forwarded to the CPU for signal detection. In order to improve achievable rate, the forwarded signal is further multiplied by a receiver filter  coefficient at the CPU. The aggregated received signal at the CPU can be written as
\begin{IEEEeqnarray}{rCl}
r_k &=&\sum_{m=1}^{M}u_{mk}\hat{g}_{mk}^*y_{m}\\&=&\sqrt{\rho}\sum_{k^{\prime}=1}^{K}\sum_{m=1}^{M}u_{mk}\hat{g}_{mk}^*g_{mk^\prime}\sqrt{q_{k^\prime}}s_{k^\prime}+
\sum_{m=1}^{M}u_{mk}\hat{g}_{mk}^*n_m.\nonumber
 \label{rk}
\end{IEEEeqnarray}
By collecting all the coefficients $u_{mk}, ~\forall ~m$ corresponding to the \textit{k}th user,  we define $\textbf{u}_k = [u_{1k}, u_{2k},\cdots, u_{Mk}]^T$ and without loss of generality, it is assumed that $\parallel \textbf{u}_k\parallel=1$. The optimal solution of $\mathbf{u}_{k}, ~q_{k},~\forall~k$ for the considered max-min SINR approach is investigated in Section IV.
\section{Performance Analysis}
In this section, we derive the average user rate for the considered system model in the previous section by following a similar approach to that in \cite{marzetta_free16}. Note that the main difference between the proposed approach and the scheme in \cite{marzetta_free16} is the new set of receiver coefficients which are introduced at the CPU to improve the achievable user rates. The benefits of the proposed approach in terms of achieved user uplink rate is demonstrated through numerical simulation results in Section V. In deriving the achievable rates of each user, it is assumed that the CPU exploits only the knowledge of channel statistics between the users and APs in detecting data from the received signal in (\ref{rk}). Without loss of generality, the aggregated received signal in (\ref{rk}) can be written as
\begin{IEEEeqnarray}{rCl}\label{rk2}
\small
r_k &&=\underbrace{\sqrt{\rho} \mathbb{E}\left\{\sum_{m=1}^Mu_{mk}\hat{g}_{mk}^*{g}_{mk}\sqrt{q_k}\right\}}_{\text{DS}_k}s_k\\
 &\!+\!&\!\underbrace{\sqrt{\rho}\left(\!\sum_{m=1}^M\!u_{mk}\hat{g}_{mk}^*\!{g}_{mk}\sqrt{q_k}\!-\!\mathbb{E}\!\left\{\!\sum_{m=1}^M\! u_{mk}\hat{g}_{mk}^*{g}_{mk}\sqrt{q_k}\!\right\}\right)}_{{\text{BU}_k}}\!s_k\nonumber\\
 &+&\sum_{k^\prime \neq k}^K\underbrace{\sqrt{\rho}\sum_{m=1}^Mu_{mk}\hat{g}_{mk}^*{g}_{mk^\prime}\sqrt{q_{k^\prime}}}_{{\text{IUI}_{kk^\prime}}}s_{k^{\prime}}
+\underbrace{\sum_{m=1}^{M}u_{mk}\hat{g}_{mk}^*n_m}_{\text{TN}_k},\nonumber
\end{IEEEeqnarray}
where $\text{DS}_k$ and $\text{BU}_k$ denote the desired signal (DS) and beamforming uncertainty (BU) for the $k$th user, respectively, and $\text{IUI}_k$ represents the inter-user-interference (IUI) caused by the $k^\prime$th user. In addition, $\text{TN}_k$ accounts for the total noise (TN) following the MRC detection. The average SINR of the received signal in (\ref{rk2}) can be defined by considering the worst-case of the uncorrelated Gaussian noise as follows \cite{marzetta_free16}:
\begin{IEEEeqnarray}{rCl}
\!\!\!\!\!{\textrm{SINR}}_k^{\text{UP}}\!=\!\dfrac{|\text{DS}_k|^2}{\!\mathbb{E}\{|\text{BU}_k|^2\!\}\!+\!\sum_{k^\prime\ne k}^K\mathbb{E}\{|\text{IUI}_{kk^\prime}|^2\!\}\!+\!\mathbb{E}\{|\text{TN}_k|^2\!\}}\!.
\label{rate}
\end{IEEEeqnarray}
\begin{figure*}[h]
\begin{IEEEeqnarray}{rCl}
\begin{split}
R_k = \log_2\left( 1+\dfrac{\textbf{u}_k^H\left(q_k\bgama_k\bgama_k^H\right)\textbf{u}_k}{\textbf{u}_k^H\left(\sum_{k^\prime\ne k}^Kq_{k^\prime}\left|\pmb{\phi}_k^H\pmb{\phi}_{k^\prime}\right|^2\bdel_{k k^\prime}\bdel_{k k^\prime}^H+\sum_{k^\prime=1}^{K}q_{k^\prime}\textbf{D}_{k k^\prime}+\dfrac{1}{\rho}\textbf{R}_{k}\right )\textbf{u}_k}\right).
\label{sinr1}
\end{split}
\end{IEEEeqnarray}
\hrulefill
\end{figure*}
Based on the SINR definition in (\ref{rate}), the achievable uplink rate of the \textit{k}th user is defined in the following theorem.
\begin{theorem}
\label{sinrf}
By employing MRC detection at APs, the achievable uplink rate of the \textit{k}th user in the Cell-free Massive MIMO system with $K$ randomly distributed single-antenna users and $M$ single-antenna APs is given by (\ref{sinr1}) (defined at the beginning of the next page). 
\end{theorem}
{Note that in (\ref{sinr1}), we have
$\bgama_k=[\gamma_{1k}, \gamma_{2k}, \cdots, \gamma_{Mk}]^T$, $\textbf{u}_k=[u_{1k}, u_{2k}, \cdots, u_{Mk}]^T$, $\bdel_{k k^\prime}=[\dfrac{\gamma_{1k}\beta_{1k^\prime}}{\beta_{1k}}, \dfrac{\gamma_{2k}\beta_{2k^\prime}}{\beta_{2k}}, \cdots, \dfrac{\gamma_{Mk}\beta_{Mk^\prime}}{\beta_{Mk}}]^T$,$\textbf{R}_{k}=\text{diag}\left [\gamma_{1k}, \gamma_{2k}, \cdots, \gamma_{Mk}\right]$, and $\textbf{D}_{kk^{\prime}}=\text{diag}\left [\beta_{1k^\prime}\gamma_{1k}, \beta_{2k^\prime}\gamma_{2k}, \cdots, \beta_{Mk^\prime}\gamma_{Mk}\right]$.
%	\begin{subequations}
%		\begin{align}
%		\label{bb1}&\bgama_k=[\gamma_{1k}, \gamma_{2k}, \cdots, \gamma_{Mk}]^T,\\
%		\label{bb2}&\textbf{u}_k=[u_{1k}, u_{2k}, \cdots, u_{Mk}]^T,\\
%		\label{bb3}&\bdel_{k k^\prime}=[\dfrac{\gamma_{1k}\beta_{1k^\prime}}{\beta_{1k}}, \dfrac{\gamma_{2k}\beta_{2k^\prime}}{\beta_{2k}}, \cdots, \dfrac{\gamma_{Mk}\beta_{Mk^\prime}}{\beta_{Mk}}]^T,\\
%		\label{bb4}&\textbf{R}_{k}=\text{diag}\left [\gamma_{1k}, \gamma_{2k}, \cdots, \gamma_{Mk}\right],\\
%		\label{bb5}&\textbf{D}_{kk^{\prime}}=\text{diag}\left [\beta_{1k^\prime}\gamma_{1k}, \beta_{2k^\prime}\gamma_{2k}, \cdots, \beta_{Mk^\prime}\gamma_{Mk}\right],
%		\end{align}
%	\end{subequations}
\\\\
{\textit{Proof:}} Please refer to Appendix A.
\section{Proposed Max-Min SINR Scheme}
\label{secop}
In this section, we formulate the umax-min SINR problem in cell-free massive MIMO, where the minimum uplink user rate between users is maximized while satisfying the transmit power constraint at each user. This max-min rate problem can be formulated as follows:
\begin{align}
P_1: \max_{q_k, \textbf{u}_k} ~~~\min_{k=1,\cdots,K} ~~R_k
\label{p1}
\end{align}
\vspace{-0.22in}
\begin{IEEEeqnarray}{rCl}
\nonumber
~~~~~~~~~~~~~~~~\text{s.t.}~~~ &&
||\textbf{u}_k||=1, ~~ \forall ~k, ~~0 \le q_k \le p_{\text{max}}^{(k)},  ~~ \forall ~k,
 \nonumber
\end{IEEEeqnarray}
where  $p_{max}^{(k)}$ is the maximum transmit power available at user \textit{k}. Problem $P_{1}$ is not jointly convex in terms of $\mathbf{u}_{k}$ and power allocation $q_{k},~ \forall~ k$. Therefore, this problem cannot be directly solved through existing convex optimization software. To tackle this non-convexity issue, we divide the original Problem $P_{1}$ into two sub-problems: receiver coefficient design (i.e. $\mathbf{u}_{k}$) and the power allocation problem. To obtain a solution for Problem $P_{1}$, these sub-problems are alternately solved as explained in the following subsections.
\subsection{Receiver Coefficients Design}
In this subsection, we solve the receiver coefficient design problem to maximize the uplink rate of each user for a given set of transmit power allocation at all users. These coefficients (i.e., $\mathbf{u}_{k}$, $\forall~k$) can be obtained by interdependently maximizing the uplink SINR of each user. Hence, the optimal coefficients for all users for a given set of transmit power allocation can be determined by solving the following optimization problem:

\begin{small}
\begin{IEEEeqnarray}{rCl}
P_2:~~~~~~~~~~~~~~~~~~~~~~~~~~~~~~~~~~~~~~~~~
~~~~~~~~~~~~~~~~~~~~~~~~~~~~~~~~~~&&\nonumber\\
\max_{\mathbf{u}_k}\!
\frac{\mathbf{u}_k^H\left(q_k\bgama_k\bgama_k^H\right)\mathbf{u}_k} {\mathbf{u}_k^H\left(\!\sum_{k^\prime\ne k}^Kq_{k^\prime}\left|\pmb{\phi}_k^H\pmb{\phi}_{k^\prime}\right|^2\!\bdel_{k k^\prime}\!\bdel_{k k^\prime}^H\!+\!\sum_{k^\prime=1}^{K}q_{k^\prime}\textbf{D}_{k k^\prime}\!+\!\frac{1}{\rho}\!\textbf{R}_{k}\right)\!\mathbf{u}_k},
 \nonumber
\label{p2}
\end{IEEEeqnarray}
\end{small}
%\vspace{-0.2in}
\begin{IEEEeqnarray}{rCl}
~\text{s.t.}~~\parallel\mathbf{u}_{k}\parallel =1,~~\forall~~k.
\end{IEEEeqnarray}

\begin{algorithm}[t]
\caption{Proposed Algorithm to Solve $P_1$}
\textbf{1.} Initialize $\textbf{q}^{(0)}=[q_1^{(0)},q_2^{(0)},\cdots,q_K^{(0)}]$, $i=1$

\textbf{2.} Repeat

\textbf{3.} $i=i+1$

\textbf{4.} Set $\textbf{q}^{(i)}=\textbf{q}^{(i-1)}$ and find the optimal receiver coefficients $\textbf{U}^{(i)}=[\textbf{u}^{(i)}_1,\textbf{u}^{(i)}_2,\cdots,\textbf{u}^{(i)}_K]$ through solving the generalized eigenvalue Problem $P_2$ in (\ref{p2})

\textbf{5.} Compute $\textbf{q}^{(i+1)}$ through solving Problem $P_4$ in (\ref{p4}).

\textbf{6.} Go back to Step 3 and repeat until required accuracy.
\label{al1}
\end{algorithm}
Problem $P_{2}$ is a generalized eigenvalue problem \cite{bookematrix}, where the optimal solutions can be obtained by determining the generalized eigenvalue of the matrix pair $\mathbf{A}_{k} = q_k\bgama_k\bgama_k^H$ and $\mathbf{B}_{k}=\!\sum_{k^\prime\ne k}^Kq_{k^\prime}|\pmb{\phi}_k^H\pmb{\phi}_{k^\prime}|^2\!\bdel_{k k^\prime}\!\bdel_{k k^\prime}^H\!+\!\sum_{k^\prime=1}^{K}q_{k^\prime}\textbf{D}_{k k^\prime}\!+\!\frac{1}{\rho}\!\textbf{R}_{k}\!$ corresponding to the maximum generalized eigenvalue.
\subsection{ Power Allocation}
In this subsection, we solve the power allocation problem for a set of fixed receiver coefficients. The power allocation problem can be formulated into the following max-min problem:
\vspace{-0.1in}
\begin{align}
P_3: \max_{q_k} ~~~\min_{k=1,\cdots,K} ~~\text{SINR}_k
\label{p3}
\end{align}
\vspace{-0.18in}
\begin{IEEEeqnarray}{rCl}
\nonumber
\text{s.t.}~~~ &&0 \le q_k \le p_{max}^{(k)},  ~~ \forall ~k, \nonumber
\end{IEEEeqnarray}
Without loss of generality, Problem $P_3$ can be rewritten by introducing a new slack variable as
 \begin{align}
P_4: ~~~\max_{t,q_k} ~~~t
\label{p4}
\end{align}
\begin{IEEEeqnarray}{rCl}
~~~\text{s.t.}~~~~~~0 &&\le q_k \le p_{max}^{(k)}, ~~ \forall~k, ~~\text{SINR}_{k}\ge t,~ \forall~k. \nonumber
\end{IEEEeqnarray}
\textit{Proposition 1}: Problem $P_{4}$ can be formulated into a GP.
\\
{\textit{Proof:}} Please refer to Appendix B.
\\
Therefore, this problem can be efficiently solved through existing convex optimization software. Based on these two sub-problems, an iterative algorithm is developed by alternately solving each sub-problem in each iteration. The proposed algorithm is summarized in Algorithm \ref{al1}.
\section{Convergence analysis}
In this section, the convergence analysis of the proposed Algorithm \ref{al1} is provided. Two sub-problems are alternately solved to determine the solution to Problem $P_1$. At each iteration, one of the design parameters is determined by solving the corresponding sub-problem while other design variable is fixed. Note that each sub-problem provides an optimal solution for the other given design variable. At the $i$th iteration, the receiver filter coefficients $\textbf{u}_{k}^{(i)},~ \forall k$ are determined for a given power allocation $\textbf{q}^{(i)}$ and similarly, the power allocation $\textbf{q}^{(i+1)}$ is updated for a given set of receiver filter coefficients $\textbf{u}_{k}^{(i)},~\forall k$. 
The optimal power allocation $\textbf{q}^{(i+1)}$
obtained
for a given $\textbf{u}_k^{(i)}$ achieves an uplink rate greater
than or equal to that of the previous iteration. In addition, the power allocation $\textbf{q}^{(i)}$ is also a feasible solution in determining $\textbf{q}^{(i+1)}$ as the receiver filter coefficients $\textbf{u}_{k}^{(i+1)},~ \forall k$ are determined for a given  $\textbf{q}^{(i)}$.  This reveals
that the achieved uplink rate monotonically increases
with each iteration, which can be also observed from the simulation
results presented in Fig. \ref{con}. As the achievable uplink max-min rate is upper bounded by a certain value for a given set of per-user power constraints, the proposed algorithm converges to a particular solution. Fortunately, the
proposed Algorithm \ref{al1} converges to the optimal solution, as we will prove by establishing the uplink-downlink duality in the following section. 
\section{Numerical Results and Discussion}
In this section, we provide numerical simulation results to validate the performance of the proposed max-min SINR approach with different parameters. A cell-free Massive MIMO system with $M$ APs and $K$ single antenna users is considered in a $D \times D$ simulation area, where both APs and users are randomly distributed. In the following subsections, we define the simulation parameters and then present the corresponding simulation results.
\begin{figure}[t!]
	\center
	\includegraphics[width=73mm]{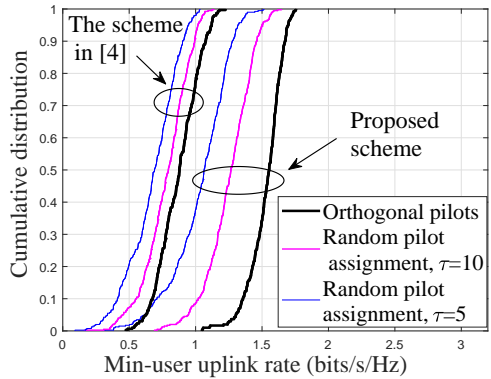}
	%\vspace{-0.25in}
	\linespread{.1}
	\caption{The cumulative distribution of the per-user uplink rate, with orthogonal and random pilots for $M=60$, $K=20$ and $D = 1$ km.}
	\label{60}
\end{figure}
To model the channel coefficients between users and APs, the coefficient $\beta_{mk}$ is given by 
$
\beta_{mk} = \text{PL}_{mk}. 10^{\frac{\sigma_{sh}z_{mk}}{10}}
\label{beta1}
$
where $\text{PL}_{mk}$ is the path loss from the $k$th user to the $m$th AP, and $10^{\frac{\sigma_{sh}~z_{mk}}{10}}$ denotes the shadow fading with standard deviation
$\sigma_{sh}$, and $z_{mk} \sim  \mathcal{N}(0,1)$ \cite{marzetta_free16}. The noise power is given by
$
P_n=\text{BW}k_BT_0W,
$
where $\text{BW}=20$ MHz denotes the bandwidth, $k_B = 1.381 \times 10^{-23}$ represents the Boltzmann constant, and $T_0 = 290$ (Kelvin) denotes the noise temperature. Moreover, $W=9$dB, and denotes the noise figure \cite{marzetta_free16}. It is assumed that that $\bar{P}_p$ and $\bar{\rho}$ denote the pilot sequence and the uplink data, respectively, where $P_p=\frac{\bar{P}_p}{P_n}$ and $\rho=\frac{\bar{\rho}}{P_n}$. In simulations, we set $\bar{P}_p=100$mW and $\bar{\rho}=100$mW.
Similar to \cite{marzetta_free16}, we suppose the simulation area is wrapped around at the edges which can simulate an area without boundaries. Hence, the square simulation area has eight neighbors. We evaluate the rate of the system over 300 random realizations of the locations of APs, users and shadowing.
\begin{figure}[t!]
\center
\includegraphics[width=73mm]{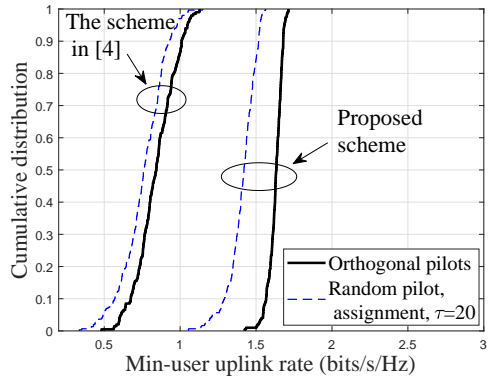}
%\vspace{-0.245in}
\linespread{.1}
\caption{The cumulative distribution of the per-user uplink rate, with random pilots for $M=100$, $K=40$, $\tau=20$, and $D=1$ km.}
\label{100}
\end{figure}
\begin{figure}[t!]
\center
\includegraphics[width=73mm]{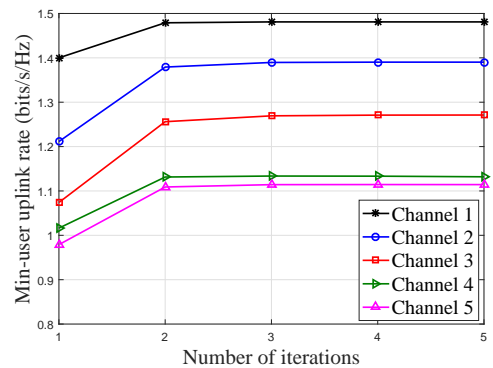}
%\vspace{-0.25in}
\linespread{.1}
\caption{The convergence of the proposed max-min SINR approach for $M=60$, $K=20$, $\tau=10$, and $D=1$ km.}
\label{con}
\end{figure}
In this subsection, we investigate the effect of the max-min SINR problem on the system performance. Fig. \ref{60} compares the cumulative distribution
of the achievable uplink rates for our proposed algorithm with the power allocation scheme in \cite{marzetta_free16}, for three cases of orthogonal pilots, random pilots with $\tau=10$ and $\tau=5$ for the length of pilot sequence. In Fig. \ref{60}, $M=60$ APs and $K=20$ users are randomly distributed through the simulation area of size $1 \times 1$ $\text{km}^2$. As the figure shows, the performance of the proposed scheme is almost three times than that of the scheme in \cite{marzetta_free16}.

In Fig. \ref{100}, we compare the performance of the proposed max-min SINR approach with the scheme in \cite{marzetta_free16} for the case of $M=100$ APs, $K=40$ users and $\tau=20$ as the length of the pilot sequence. 
Fig. \ref{100} shows the superiority of the proposed iterative algorithm over the power allocation scheme in \cite{marzetta_free16}. Moreover, Fig. \ref{100} demonstrates that the rate of the proposed max-min SINR approach is more concentrated around the median.
Fig. \ref{con} investigates the convergence of the
proposed max-min SINR algorithm for a set of different channel realizations.
The figure shows that the proposed algorithm
converges after a few iterations, while the minimum rate of the users increases with
the iteration number.
\vspace{-.1cm}
\section{Conclusions}
We have considered the max-min optimization problem in cell-free Massive MIMO systems, and propose an efficient solution that maximizes the smallest of the uplink rate of the users. We propose to divide the original max-min problem into two sub-problems which can be iteratively solved by exploiting generalized eigenvalue problem and GP. The simulation results showcased the effectiveness of the proposed scheme in terms of maximising the smallest of the uplink rate of the users compared with existing schemes.
\vspace{-.1cm}
\section*{Appendix A: Proof of Theorem \ref{sinrf}}
The desired signal for the user $k$ is given by
\vspace{-.1cm}
\begin{IEEEeqnarray}{rCl}
 \small
\text{DS}_k\!=\! \mathbb{E}\left \{\!\sum_{m=1}^{M}\!u_{mk}\hat{g}_{mk}^*g_{mk}\!\sqrt{q_k}\right \}\!=\!\sqrt{q_k}\!\sum_{m=1}^{M}\!u_{mk}\!\gamma_{mk}.~
\end{IEEEeqnarray}
Hence,
$
 \left|\text{DS}_k \right| ^2 = q_k \left(\sum_{m=1}^{M}u_{mk}\gamma_{mk}\right)^2.
\label{dsk}
$
Moreover, the term $\mathbb{E}\{\left | \text{BU}_k\right |^2\}$ can be obtained as
\begin{IEEEeqnarray}{rCl}
\mathbb{E}\!\!\!\!\!\!&&\left\{ \left | \text{BU}_k \right | ^2\right\} 
=\rho \mathbb{E}  \left
\{ \left| \sum_{m=1}^Mu_{mk}\hat{g}_{mk}^*{g}_{mk}\sqrt{q_k}\right.\right.\\
 &-&\left.\left.\rho\mathbb{E}\left\{\sum_{m=1}^Mu_{mk}\hat{g}_{mk}^*{g}_{mk}\sqrt{q_k}\right\}  \right|^2 \right \} =\rho q_k \sum_{m=1}^Mu_{mk}^2\gamma_{mk}\beta_{mk},\nonumber
\label{ebuk}
\end{IEEEeqnarray}
where the last equality comes from the analysis in \cite[Appendix A]{marzetta_free16}] and using the following fact; $\gamma_{mk}=\mathbb{E}\{\left |\hat{g}_{mk}\right | ^2\}=\sqrt{\tau p_p}\beta_{mk}c_{mk}$. The term $\mathbb{E}\{\left | \text{IUI}_{k k^\prime}\right |^2\}$ is obtained as
\begin{IEEEeqnarray}{rCl} \label{eiui}
\mathbb{E}& \{| &\text{IUI}_{k k^\prime} |^2 \}= \rho\mathbb{E} \left\{ \left | \sum_{m=1}^Mu_{mk}\hat{g}_{mk}^*{g}_{mk^\prime}\sqrt{q_{k^\prime}}\right |^2\right\}\nonumber\\
 &=&\rho\mathbb{E} \left\{\left| \sum_{m=1}^Mc_{mk}u_{mk}{g}_{mk^\prime}\sqrt{q_{k^\prime}}\right.\right.\nonumber\\
 &&\left.\left.\left(\sqrt{\tau p_p}\sum_{i=1}^{K}g_{mi}\pmb{\phi}_k^H\pmb{\phi}_i\!+\!\pmb{\phi}_k^H\textbf{n}_{p,m}\right)^*\right|^2\right\}\nonumber\\
&=& \rho \underbrace{q_{k^\prime} \mathbb{E}\left \{\left |\sum_{m=1}^Mc_{mk}u_{mk}g_{mk^\prime}\tilde{n}_{mk}^*  \right |^2\right\}}_{A}
\\
 &\!+\!&\rho \underbrace{\tau p_p\mathbb{E}\! \left \{ \!q_{k^\prime}\left | \! \sum_{m=1}^M\!c_{mk}u_{mk}{g}_{mk^\prime}\left(\sum_{i=1}^{K}g_{mi}\!\pmb{\phi}_k^H\pmb{\phi}_i\!\right)^*\!\right |^2 \right \} }_{B},\nonumber
\end{IEEEeqnarray}
where the third equality in (\ref{eiui}) is due to the fact that for two independent random variables $X$ and $Y$ and $\mathbb{E}\{X\}=0$, we have $\mathbb{E}\{\left | X+Y \right |^2\}=\mathbb{E}\{\left | X \right |^2\}+\mathbb{E}\{\left | Y \right |^2\}$ \cite[Appendix A]{marzetta_free16}.
Since $\tilde{n}_{mk}=\pmb{\phi}_k^H\textbf{n}_{p,m}\sim \mathcal{CN}(0,1)$ is independent from the term $g_{mk^\prime}$ similar to \cite[Appendix A]{marzetta_free16}, the term $A$ in (\ref{eiui}) immediately is given by
$
A = q_{k^\prime} \sum_{m=1}^Mc_{mk}^2u_{mk}^2\beta_{mk^\prime}.
$The term $B$ in (\ref{eiui}) can be obtained as
\begin{IEEEeqnarray}{rCl} \label{b1}
%\vspace{-0.04in}
\small
B &=&  \underbrace{\tau p_p q_{k^\prime}\mathbb{E}\left \{\left | \sum_{m=1}^M c_{mk}u_{mk}\left|{g}_{mk^\prime}\right|^2\pmb{\phi}_k^H{\pmb{\phi}}_{k^\prime}\right|^2\right\}}_{C}\\
 &+&\underbrace{\tau p_p q_{k^\prime}\mathbb{E} \left\{\left| \sum_{m=1}^M c_{mk}u_{mk} {g}_{mk^\prime} \left(\sum_{i\ne k^\prime}^{K}g_{mi}\pmb{\phi}_k^H\pmb{\phi}_i\right)^* \right |^2 \right\}}_{D}.\nonumber
\end{IEEEeqnarray}
The first term in (\ref{b1}) is given by
% \vspace{-0.07in}
\begin{IEEEeqnarray}{rCl}
\small
C &=& \tau p_p q_{k^\prime}\mathbb{E}\left \{\left |\sum_{m=1}^M c_{mk}u_{mk}\left |{g}_{mk^\prime}\right |^2\pmb{\phi}_k^H{\pmb{\phi}}_{k^\prime} \right| ^2 \right\}\nonumber\\
 &=&\tau p_p q_{k^\prime}\left |\pmb{\phi}_k^H{\pmb{\phi}}_{k^\prime}\right |^2\sum_{m=1}^Mc_{mk}^2u_{mk}^2\beta_{mk^\prime}^2\nonumber\\
 &+&
 q_{k^\prime}\left |\pmb{\phi}_k^H{\pmb{\phi}}_{k^\prime}\right |^2\left(\sum_{m=1}^M u_{mk}\gamma_{mk}\dfrac{\beta_{mk^\prime}}{\beta_{mk}}\right)^2,
\end{IEEEeqnarray}
where the last equality is derived based on the fact $\gamma_{mk}=\sqrt{\tau p_p}\beta_{mk}c_{mk}$. The second term in (\ref{b1}) can be obtained as
% \vspace{-0.09in}
\begin{IEEEeqnarray}{rCl}
D &=& \tau p_p q_{k^\prime}\mathbb{E}\left \{\left | \sum_{m=1}^Mc_{mk}u_{mk} {g}_{mk^\prime} \left(\sum_{i\ne k^\prime}^{K}g_{mi}\pmb{\phi}_k^H\pmb{\phi}_i\right)^* \right |^2 \right\}\nonumber\\
 &=&
 \tau p_p\sum_{m=1}^{M}\sum_{i \ne k^\prime}^{K} q_{k^\prime} c_{mk}^2u_{mk}^2\beta_{mk^\prime}\beta_{mi}\left |\pmb{\phi}_k^H\pmb{\phi}_{i}\right |^2.
 \label{d}
\end{IEEEeqnarray}
% \vspace{-0.03in}
Hence, (\ref{eiui}) can be written as
\begin{IEEEeqnarray}{rCl}
\small
\mathbb{E}\!\!\!\!\!\!&&\left\{\left| \text{IUI}_{k k^\prime}\right|^2\right\}= \underbrace{q_{k^\prime} \sum_{m=1}^Mc_{mk}^2u_{mk}^2\beta_{mk^\prime}}_{C_1}~~~~~~~~~~~~~\\
 &+&\tau p_p q_{k^\prime}\left |\pmb{\phi}_k^H{\pmb{\phi}}_{k^\prime}\right |^2
 \sum_{m=1}^Mc_{mk}^2u_{mk}^2\beta_{mk^\prime}^2
\nonumber\\
 &+&\tau p_pq_{k^\prime}\sum_{m=1}^{M}\sum_{i \ne k^\prime}^{K}c_{mk}^2u_{mk}^2\beta_{mk^\prime}\beta_{mi}\left| \pmb{\phi}_k^H\pmb{\phi}_{i}\right|^2\nonumber\\
 &+& q_{k^\prime}\left |\pmb{\phi}_k^H{\pmb{\phi}}_{k^\prime}\right |^2\left(\sum_{m=1}^M u_{mk}\gamma_{mk}\dfrac{\beta_{mk^\prime}}{\beta_{mk}}\right)^2,
\end{IEEEeqnarray}
and
\begin{IEEEeqnarray}{rCl}
C_2&=& \tau p_p q_{k^\prime}\left|\pmb{\phi}_k^H{\pmb{\phi}}_{k^\prime}\right|^2
 \sum_{m=1}^Mc_{mk}^2u_{mk}^2\beta_{mk^\prime}^2\nonumber\\
 &+&
\underbrace{\tau p_p q_{k^\prime}\sum_{m=1}^{M}\sum_{i \ne k^\prime}^{M} c_{mk}^2u_{mk}^2\beta_{mk^\prime}\beta_{mi}\left|\pmb{\phi}_k^H\pmb{\phi}_{i}\right|^2}_{C_3}.
\label{c2}
\end{IEEEeqnarray}
For the last term of (\ref{c2}), we have
\begin{IEEEeqnarray}{rCl}
C_3&=& \tau p_p q_{k^\prime}\sum_{m=1}^{M}\sum_{i \ne k^\prime}^{K} c_{mk}^2u_{mk}^2\beta_{mk^\prime}\beta_{mi}\left| \pmb{\phi}_k^H\pmb{\phi}_{i}\right|^2\nonumber\\
 &=&\sqrt{\tau p_p}q_{k^\prime}\sum_{m=1}^{M}u_{mk}^2c_{mk}\beta_{mk^\prime}\beta_{mk}-q_{k^\prime}\sum_{m=1}^{M}u_{mk}^2c_{mk}^2\beta_{mk^\prime}\nonumber\\
 &-&\tau p_p q_{k^\prime}\sum_{m=1}^{M}u_{mk}^2c_{mk}^2\beta_{mk^\prime}\left| \pmb{\phi}_k^H{\pmb{\phi}}_{k^\prime}\right|^2, 
\end{IEEEeqnarray}
where in the last step, we used equation (\ref{cmk}). As a result, $C_1+C_2=\sqrt{\tau p_p}q_{k^\prime}\sum_{m=1}^{M}u_{mk}^2c_{mk}\beta_{mk^\prime}\beta_{mk}$. Then finally we have
\begin{IEEEeqnarray}{rCl}
\mathbb{E}\left\{\left|\text{IUI}_{k k^\prime}\right|^2\right\}&&=\rho q_{k^\prime}\left(\sum_{m=1}^{M}u_{mk}^2\beta_{mk^\prime}\gamma_{mk}\right)\\
 &+&\rho q_{k^\prime} \left|\pmb{\phi}_k^H{\pmb{\phi}}_{k^\prime}\right|^2 \left(\sum_{m=1}^{M}u_{mk} \gamma_{mk}\dfrac{\beta_{mk^\prime}}{\beta_{mk}}\right)^2.\nonumber
 \label{euiu}
\end{IEEEeqnarray}
The total noise for the user $k$ is given by
\begin{IEEEeqnarray}{rCl}
\!\!\!\!\!\!\!\!\!\mathbb{E}\left\{\left|\text{TN}_k\right|^2\right\}\!=\!\mathbb{E}\left\{\left|\sum_{m=1}^{M}u_{mk}\hat{g}_{mk}^*n_m \right|^2\!\right\}\!=\!\sum_{m=1}^{M}u_{mk}^2\gamma_{mk},
\label{tn}
\end{IEEEeqnarray}
where the last equality is due to the fact that the terms $\hat{g}_{mk}$ and $n_m$ are uncorrelated.
Finally, by substituting (\ref{dsk}), (\ref{ebuk}), (\ref{euiu}) and (\ref{tn}) into (\ref{rate}), SINR of $k$th user is obtained by (\ref{sinr1}), which completes the proof of Theorem 1.
~~~~~~~~~~~~~~~~~~~~~~~~~~~~~~$\blacksquare$
\vspace{-.1cm}
\section*{Appendix B: Proof of Proposition 1}
The standard form of GP is defined as follows \cite{bookboyd}:
\begin{align}
P_5: \min~~~f_0(\textbf{x})
\label{p5}
\end{align}
\vspace{-0.3in}
\begin{IEEEeqnarray}{rCl}
~~~\text{s.t.}~~~~~~ && f_i(\textbf{x}) \le 1, ~~i= 1,\cdots, m,
\nonumber\\
~~~~~~~~~~~~~~~~~~~~~~  && g_i(\textbf{x}) = 1, ~~i= 1,\cdots, p,
\nonumber
\end{IEEEeqnarray}
where $f_0$ and $f_i$ are posynomial and $g_i$ are monomial functions. Moreover, $\textbf{x}=\{x_1,\cdots,x_n\}$ represent the optimization variables. The SINR constraint in  (\ref{p4}) is not a posynomial function in its form, however it can be rewritten into the following posynomial function:
\begin{IEEEeqnarray}{rCl}
\small
\dfrac{\mathbf{u}_k^H\!\left(\!\sum_{k^\prime\ne k}^K\!q_{k^\prime}\!|\!\pmb{\phi}_k^H\!\pmb{\phi}_{k^\prime}|^2\!\bdel_{k k^\prime}\!\bdel_{k k^\prime}^H\!+\!\sum_{k^\prime=1}^{K}\!q_{k^\prime}\!\textbf{D}_{k k^\prime}\!+\!\frac{1}{p}\!\textbf{R}_{k}\!\right)\!\mathbf{u}_k}{\mathbf{u}_k^H\!\left(q_k\bgama_k\bgama_k^H\right)\mathbf{u}_k}\!&<&\!\dfrac{1}{t},\nonumber \\  \forall k.
\label{inv}
\end{IEEEeqnarray}
By applying a simple transformation, (\ref{inv}) is equivalent to the following inequality:
\begin{IEEEeqnarray}{rCl}
q_k^{-1}\left(\sum_{k^\prime\ne k}^K\!a_{kk^\prime}q_{k^\prime}\!+\!\sum_{k^\prime=1}^{K}b_{kk^\prime}q_{k^\prime}+c_k\right)<\dfrac{1}{t},
\label{inv2}
\end{IEEEeqnarray}where
$a_{kk^\prime}=\frac{\mathbf{u}_k^H (|\!\pmb{\phi}_k^H\!\pmb{\phi}_{k^\prime}|^2\!\bdel_{k k^\prime}\!\bdel_{k k^\prime}^H) \mathbf{u}_k}{\mathbf{u}_k^H\!(\bgama_k\bgama_k^H)\mathbf{u}_k}$,
$b_{kk^\prime}=\frac{\mathbf{u}_k^H \textbf{D}_{k k^\prime}\mathbf{u}_k}{\mathbf{u}_k^H\!(\bgama_k\bgama_k^H)\mathbf{u}_k}$, and
$c_k=\frac{\mathbf{u}_k^H \textbf{R}_{k} \mathbf{u}_k}{p\mathbf{u}_k^H\!(\bgama_k\bgama_k^H)\mathbf{u}_k}$.
The transformation in (\ref{inv2}) shows that the left-hand side of (\ref{inv}) is a polynomial function. Therefore, the power allocation Problem $P_4$ is a standard GP (convex problem), where the objective function and constraints are monomial and polynomials, which completes the proof of Proposition 1. ~~~~~~~~~~~~~~~~~~~~~~~~~~~~~~~~~~
~~~~~~~~~~~~~~~~~~~~~~~~~~~~~~~~~~~$\blacksquare$
%where $a_{kk^\prime}=\frac{\mathbf{u}_k^H (|\!\pmb{\phi}_k^H\!\pmb{\phi}_{k^\prime}|^2\!\bdel_{k k^\prime}\!\bdel_{k k^\prime}^H) \mathbf{u}_k}{\mathbf{u}_k^H\!(\bgama_k\bgama_k^H)\mathbf{u}_k}$, $b_{kk^\prime}=\frac{\mathbf{u}_k^H \textbf{D}_{k k^\prime}\mathbf{u}_k}{\mathbf{u}_k^H\!(\bgama_k\bgama_k^H)\mathbf{u}_k}$, and finally $c_k=\frac{\mathbf{u}_k^H \textbf{R}_{k} \mathbf{u}_k}{p\mathbf{u}_k^H\!(\bgama_k\bgama_k^H)\mathbf{u}_k}$.
%\section{Acknowledgement}
%We would like to thank Dr. Hien Quoc Ngo, postdoctoral researcher of the Division for Communication Systems in the Department of Electrical Engineering (ISY) at Linkoping University, Sweden, for his valuable guidance and the useful discussion.
\bibliographystyle{IEEEtran}
\bibliography{conf_TWC_cameraready} 
\end{document}